# Searching for Life on Habitable Planets and Moons

**Ashwini Kumar Lal**

Ministry of Statistics & Programme Implementation
New Delhi, India
ashwinikumar_lal@yahoo.co.in

**Abstract**

Earth is the only known inhabited planet in the universe to date. However, advancements in the fields of astrobiology and observational astronomy, as also discovery of large varieties of extremophiles with extraordinary capability to thrive in the harshest environments on Earth, have led to speculation that life may possibly be thriving on many of the extraterrestrial bodies in the universe. Coupled with the growing number of exoplanets detected over the past decade, the search for the possibility of life on other planets and satellites within the solar system and beyond has become a passion as well as a challenge for scientists in a variety of fields. This paper examines such possibility in the light of findings of the numerous space probes and theoretical research undertaken in this field over the past few decades.

## 1. Introduction

Many scientists and religious leaders believe life originated on Earth through processes that gave life to non-life. Many respond skeptically to the possibility there may be life on other planets. In contrast to the widely accepted beliefs in abiogenesis where the Earth is seen as the centre of the biological universe, theories of panspermia view life as widespread throughout the cosmos. Although panspermia does not address the fundamental question as to when and where exactly life originated first in the universe, it nonetheless does lend support to the possibility of life being found on other planets (besides Earth) and satellites within the solar system and beyond (Lal 2008).

Central to theories of panspermia is the belief that the 'seeds of life' are ubiquitous (Arrhenius 1908, 2009), and that these 'seeds of life' i.e. living creatures were embedded in meteors, asteroids and comets, and deposited upon Earth as well as to other habitable bodies in the universe (Wickramasinghe et al. 2009 ; Joseph 2009). Considerable evidence has been presented in support of pansermia by Hoyle and Wickramasinghe (Hoyle and Wickramasinghe 1977, 2000 ; Wickramasinghe 1995). Wickramasinghe and colleagues have provided nearly convincing evidence that dust grains in interstellar clouds could contain spores, desiccated bacteria, and living microbes which survive in comets on cloud condensation to stars and planets.

The large-scale presence of organic molecules in the interstellar clouds, comets and asteroids, and evidence of amino acids in carbonaceous meteorites support a cosmic perspective on the origin of life. One hundred fifty different chemical compounds including several organic compounds and amino acids with C, H, O, and N as major constituents have been detected in the interstellar clouds, circumstellar envelopes, and comets since 1965 (Lammer et al. 2009). Biomarkers like glycine ($CH_2NH_2COOH$)- simplest amino acid and glycolaldehyde ($CH_2OHCHO$) - simplest sugar have been identified in Sagittarius B2, a dense star-forming cloud of interstellar gas at the heart of Milky Way Galaxy (Sorrell 2001; Kuan et al. 2003).

Mars, the Jovian satellites - Europa, Ganymade, Io, and Callisto; and Saturn's Enceladus and Titan are already hotspots for search for life within the solar system (Naganuma and Sekine 2009; Schulze- Makuch 2009). Search for life beyond the solar system is currently underway with the launching of NASA's Kepler (March 2009) and French CNES-ESA's CoRoT(December 2006) Missions. While the former has been designed to find Earth-size planets in the habitable zone of sun-like stars, the latter has been looking for Earth-like planets around nearby stars in our parent galaxy. The Kepler Mission has been continuously and simultaneously observing 150,000 stars using transit method of detection for exoplanets wherein astronomers determine size (radius) of an exoplanet transiting in front of its parent star's disk through observed dip in the visual brightness of the stars. Kepler's highly sensitive photometric camera has already measured hundreds of possible planetary signatures that are currently being analyzed (NASA 2010).

In 1992, Wolszczan and Frail first reported in *Nature* about the discovery of a planetary system (PSR 1257+12B and PSR 1257+12C) around a millisecond pulsar (PSR 1257+12) outside our solar system in the constellation Virgo. Detection of the first known extrasolar planet, '51 Pegasi b' orbiting a main sequence star, using radial velocity detection technique, was announced 3 years later in 1995. The planet was found to be in a four-day orbit around the nearby G-type star, '51 Pegasi' with a tiny semi- major axis of 0.05 AU - several times closer to its parent star than Mercury is to our Sun (Mayor and Queloz 1995). Till May 2013, Around 900 exoplanets of three distinct types - gas giants (hot Jupiters), hot super-Earths in short period orbits, and ice giants have been discovered beyond our solar system. Mayor et al. (2009) have reported discovery of an Earth-like exoplanet, 'GJ 581e' with mass only 1.9 times Earth's mass at an orbital distance of just 0.03 AU from its parent star. The planet orbits its parent star once every 3.15 days. The star, "Gliese 581", around which the exoplanet revolves, falls into the category of low-mass spectral type-M red dwarf stars, around which low-mass planets in the habitable zones are most likely to be found. A number of planets have been discovered orbiting red dwarf stars. Red dwarfs constitute about 75% of the starts closest to our solar system. The majority of the discovered exoplanets to date, however, belong to the category of gas giants, where chances of finding life are almost ruled out. Guo et al. (2009) have estimated that the number of terrestrial planets in the habitable zones of host stars in the Milky Way with masses in the range of 0.08 - 4Msun is 45.5 billion. The majority of terrestrial planets in habitable zones in the Milky Way possibly orbit K- type stars.

## 2. Circumstellar Habitable Zone

Circumstellar Habitable Zone (CHZ), also known as 'Godilocks Zone', is the region around a star within which an Earth-like planet can sustain liquid water on its surface, a condition necessary for photosynthesis and supporting life (Fig.1&2). Astronomers in the 1970's referred to Goldilocks Zone as a remarkable small region of space that is neither too hot, nor too cold, and is just right for sustaining life. However, in the 1970's, it was thought that life could not survive under extreme conditions i.e. no colder than Antarctica, no hotter than scalding waters, no higher than clouds, and no lower than a few miles.

However, discoveries over the past four decades, call for revision of the classical definition of Goldilocks Zone. These include the discovery of the whole ecosystems around deep-sea vents where sunlight never penetrates, and deep sea thermal vents where water is hot enough to melt lead, and the discovery of extremophiles with extraordinary capability to survive and thrive in the harshest environments on Earth intolerable to virtually all other living creatures - scalding waters, subzero temperatures, extreme radiation, bone-crushing pressures, corrosive acidic, and extreme salty and alkaline conditions.

The concept of 'Circumstellar Habitable Zone'(also referred to as 'ecosphere') was first proposed by Huang (1960). The circumstellar habitable zone (CHZ) is a spherical shell around a main sequence star where an atmosphere can support liquid water at a given time. Liquid water is an important requirement for habitability because of its essential role as a solvent in biochemical reactions. It is perceived as the best solvent for carbon-based life to emerge and evolve thereafter. It is an ideal mixing bowl to to create increasingly complex molecules.

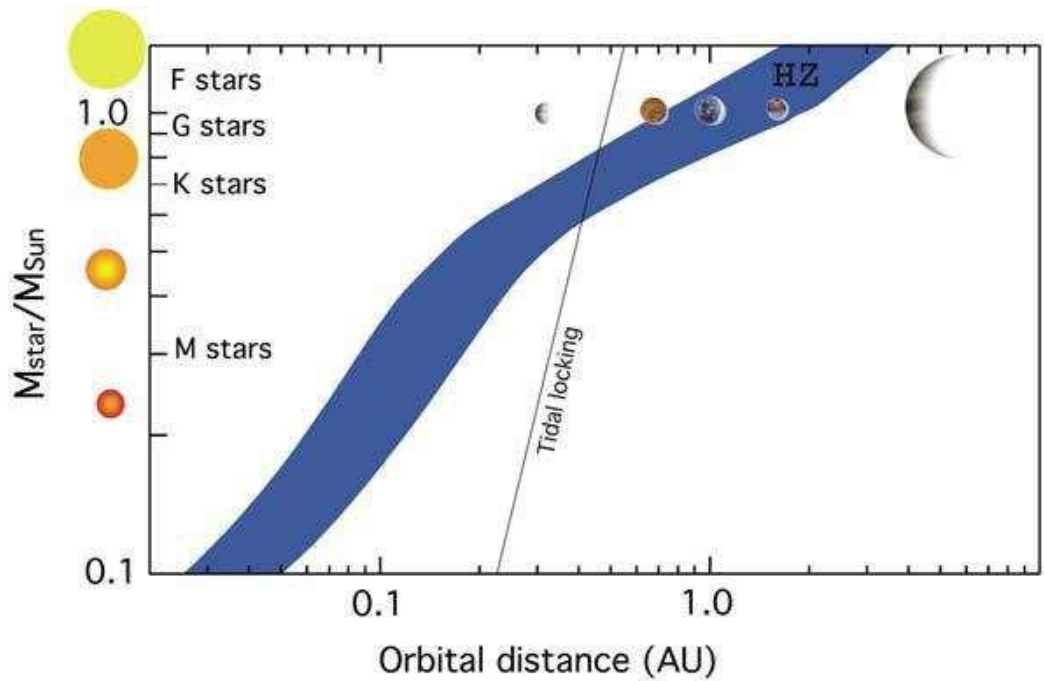

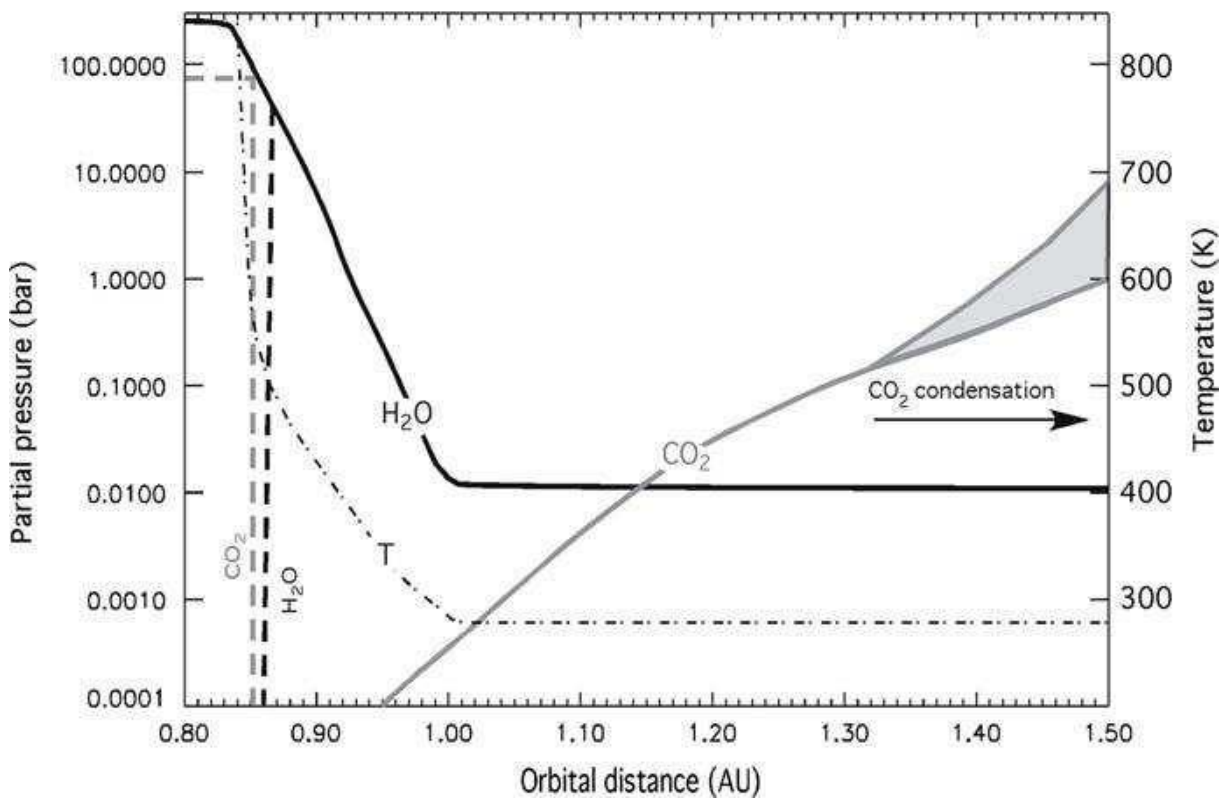

**Fig. 1&2** The HZ (*upper panel*) and the chemistry composition (*lower panel*) of an Earth-analog atmosphere as a function of distance from its host star. The *dashed-dotted line* represents the surface temperature of the planet and the *dashed lines* correspond to the inner edge of the HZ where the greenhouse conditions vaporize the whole water reservoir (**adapted from Kaltenegger and Selsis 2007**)

Habitable zones are bounded by the range of distances from a star for which liquid water can exist on a planetary surface, depending on such additional factors as the nature and density of its atmosphere, and its surface gravity. For a G-type star such as our Sun, the traditional Habitable Zone lies between 0.95 AU and 1.37 AU (Kasting et al. 1993). The limits of the Sun's HZ are, however, not easy to fix. Discovery of primitive life, or fossils on Mars would extend the HZ of the Sun out to at least 225 million km (~ 1.5 AU). Indeed, Mars itself is believed to have enjoyed warmer, wetter conditions in its early history, under which primitive organisms may have evolved.

Much farther from Sun are the gas giants and their larger moons - at first sight inhospitable to life. The effect of tidal heating on those worlds orbiting the gas giants, e.g. Europa and Callisto, leading to existence of subsurface liquid water, makes them biologically interesting (Reynolds et al. 1987). If life is eventually discovered on any of the Jupiter's moons, the outer margin of the Sun's habitable zone would be pushed out to almost 800 million km (~ 5.33 AU).

The hot inner edge of a CHZ is located at the orbital distance where a planet's water is broken up by stellar radiation into oxygen and hydrogen. It is believed that massive disassociation of planetary water occurred on Venus (which has an average orbital distance of 0.7AU) in the remote past of the planet's history via runaway greenhouse effect. Conventionally, if a planet is located outside the habitable zone, it is believed to be incapable of sustaining life since it is impossible for liquid water to exist beyond the band. The size and location of CHZs of other stars depend on various factors such as size, mass, brightness, and temperature of the star, and other planetary factors such as knowledge of a habitable planet's orbit including its eccentricity.

Some astronomers are critical of the classical habitable zone theory as it takes into consideration only carbon-based life. Moreover, the classical HZ is defined for surface conditions only. Chemolithotrophic life, a metabolism that does not depend on the stellar light, can still exist outside the HZ thriving in the interiors of the planet where liquid water is available. Such metabolisms do not depend on photosynthesis for their energy requirement and growth. Many extremophiles have been discovered to be thriving deep within Earth's crust and hydrothermal vents in deep ocean along the mid - oceanic ridges, such as East Pacific Ridge and Mid-Atlantic Ridge. Psychrophilic and psychrotrophic (cold-loving) microbes that inhabit permanently frozen regions of permafrost, polar ice-sheets, and glaciers, and deep ocean and deep-sea sediments on Earth provide analogues for microbial life that might inhabit ice sheets and permafrost of Mars, comets, or the ice/water interfaces or sediments deep beneath the icy crust of Europa, Callisto, or Ganymede (Pikuta and Hoover 2003).

Mass of a planet is crucial to its habitability. Planets much smaller than 1 Mearth may lose atmospheric gases including H and He. Mars (at ~0.1 Mearth) is a good example (Lammer et al. 2008). Smaller planets tend to lose their internal heat faster, thereby removing the energy source required to drive plate tectonics. Recycling of carbonate rocks to refresh atmospheric $CO_2$ is thought to be a key factor in stabilizing Earth's long-term climate (Kasting et al. 1993). At the other end of the mass scale, planets larger than about 10 Mearth are considered likely to capture nebular gas during their formation and evolve into gas-or ice-giants (Valencia et al. 2007).

The range of star types that can support Earth-like life on planets may be limited to those lower mass stars that live long enough as stable luminous stars for planets to form and complex life to evolve. Jones et al. (2006) have estimated that it takes at least 1 Gyr to evolve life on a habitable planet.

Potentially habitable planets around lower mass stars such as M-type red dwarfs must orbit closer to these stars - perhaps one-fiftieth the distance of Earth to the Sun. This is because these stars are small and generate less heat than our Sun. Although all main sequence stars generate luminous energy by converting hydrogen into helium through thermonuclear fusion, stars more massive than 1.5 times that of the Sun (i.e. stars of spectral types - O, B, or A dwarfs like Sirius) age too quickly to support the development of complex Earth-type life. Even the largest, possibly suitable stars-spectral type FO-4 may only be able to support Earth-type life for only around 2 billion years. Planets in favourable orbits around such stars thus may not have sufficient time to develop complex life such as plants and animals (Kasting et al. 1993). Moreover, based on traditional views of star and planet formation, within a couple of billion years of a star's birth, cometary and asteriodal bombardment may still be so intense that surviving on such planets would be nearly impossible.

On the opposite extreme, life may be unable to develop on planets orbiting stars with less than half of solar mass (e.g., smaller spectral type - M dwarfs like Porxima Centauri) These stars are more likely to tidally lock planets into such proximal orbits that any surface liquid water will evaporate too quickly, before life can develop (Peale 1977). Tidal locking (synchronous rotation of the star and the planet) may eventually cause the destruction of any life including those which might rain down upon a planet due to panspermia. These planets would be unable to develop sustaining atmosphere through condensation on the cold, perpetually dark side of the planet. Most M-type red dwarfs stars regularly emit larger stellar flares which would tend to sterilize life on a close orbiting Earth-type planet.

The large majority of stars close to our Sun fall in the lower mass domain (M, K). A star, with 25% of the luminosity of the Sun, will have a CHZ centred at about 0.50 AU. In contrast, a star twice the Sun's luminosity will have a CHZ centred at about 1.4 AU. Low mass stars have closer orbital locations of their HZs (0.02AU-0.70 AU). They are longer active in X-rays and EUV radiation. Earth-like exoplanets may be exposed to much stronger by coronal mass ejections (CMEs) and stellar winds than at sun-like stars within 1 AU.

Majority of the planets with different masses found with semi-major axis of ~0.03 AU are around the host stars with masses ~0.10 $M_{earth}$ in the terrestrial planets (Ida and Lin 2005).Incidentally, width of the HZs of such stars is 0.03 AU. This gives rise to the probability of presence of large number of terrestrial planets in the HZs around these stars in the Milky Way being very high.

NASA's Kepler mission is currently searching for habitable Earth-size planets around nearby main sequence stars that are less massive than spectral type A, but more massive than M-dwarf stars of types - F,G, and K. Over 65 percent of the main sequence stars in the solar neighbourhood, that are possibly suitable (i.e, with a stellar mass between 0.5 and 1.5 times that of Sun) for hosting Earth-type planets, may be members of binary or multiple star systems (Duquennoy and Mayor 1991). Among the currently known extrasolar planet-hosting stars, ~25% are members of binaries (Haghighipour 2006). In binary star systems, however, a planet must not be located too close to 'home star' or its orbit will be unstable making it impossible for complex life to evolve.

In star systems with more than two stars, the limits on stable orbital distance are so stringent that the presence of Earth-like planets, where surface water would be liquid, is much less likely. In March 2007, astronomers using NASA's Infrared Spitzer Space Telescope, announced their findings that planetary systems and dusty disk of asteroids, comets, and perhaps planets may be at least as abundant in binary systems as they are around single stars.

## 2.1. Computation of Habitable Zone

If the habitable zone is defined simply as the distance of a star where the effective temperature is in the range of 0°C to 100°C, it is then straightforward to calculate the radii of HZ's inner and outer bounds. The relevant formula in this connection is :

$$L = 4 \pi r^2 \sigma T^4 \quad \text{......................................... (i)}$$

where, L = star's luminosity,

r = the distance from the centre of star,

σ = Stefan - Boltzmann constant,

= $5.67 \times 10^{-8}$ W m$^{-2}$ K$^{-1}$, and

T = effective temperature (in Kelvin)

For the Sun, equation (i) yields a habitable range for the HZ of 0.7 to 1.5 AU. The HZ range for other stars can then be calculated from the formula given below :

$$L(\text{star})/L(\text{sun}) = r(\text{star})^2 / r(\text{sun})^2 \quad \text{........................ (ii)}$$

In the case of Vega, L (star)/ L (sun) = 53, which gives a range of HZ of 5.1 to 10.9 AU.

## 3. Galactic Habitable Zone (GHZ)

The Galactic Habitable Zone (GHZ) is a hypothesized spherical band that may be a likely place for terrestrial life to develop in a galaxy. It is essentially extension of circumstellar habitable zone to a galactic scale. In the present context, GHZ refers to habitable zone in our parent galaxy, the Milky Way.

Lineweaver et al. (2004) have suggested four prerequisites for traces of life to be found on exoplanets within the Milky Way Galaxy : the presence of a host star, enough heavy elements to form terrestrial planets, sufficient time for biological evolution, and an environment free from life - extinguishing supernovae. They have defined the Galactic Habitable Zone as an annular region between 7 to 9 kiloparsecs (23,000 to 29,000 light years) from the galactic centres, and is composed of stars that formed between 8 to 4 billion years ago.

According to planetary astronomer, Guillermo Gonzalez (2005), the width of GHZ is controlled by two factors. The inner (closest to the center of the galaxy) limit is set by threats that permit development of complex life: nearby transient sources of ionizing radiation and comet impacts. Such threats tend to increase close to the galactic center. The outer limit is imposed by galactic chemical evolution, specifically the abundance of heavier elements. Observation of stars in the Milky Way Galaxy suggests that the outer reaches of a spiral galaxy may be too poor in heavy elements to allow terrestrial complex life to exist.

Based on current theory and studies of extrasolar planets, metal-rich stars are more likely to have planets orbiting around them, as a certain minimum amount of metals is needed to form rocky bodies. A metallicity of at least half that of the Sun is required to build a habitable terrestrial planet. The mass of a terrestrial planet has important consequences for interior heat loss, volatile inventory, and loss of atmosphere (Gonzalez et al. 2001). According to Valencia et al. (2006), terrestrial planets are rocky planets from one to ten Earth masses with the same chemical and mineral composition as the Earth.

Presence of heavy elements is a major consideration in demarcation of GHZ. The rate at which massive stars form drops sharply as one ventures outward from the Milky Way's centre, and the abundance of heavy elements fall with them (Lineweaver et al. 2004). According to one calculation, when the Sun formed 4.6 Gya, the outer third of the galaxy lacked enough heavy elements to support life. Elements have since become more widely distributed. Now, only the galaxy's outer rim is too undernourished to form Earth-like planets easily. The classical definition of GHZ excludes stars too close to the galactic centre, if they are metal-rich star. This is exactly the case with our Sun which is conveniently distanced about 28,000 light years from the galactic centre.

Location on the outside of or outskirts of the galaxy's spiral arms and maintaining an orbital speed which prevents the star from being swept up and into the spiral arm is another requirement of the GHZ. Our Sun revolves at the same rate as the galaxy's spiral arm rotation. This synchronization and the unusually circular orbit of our Sun around the galactic centre keeps it clear of the spiral arms where numerous stars and other life-neutralizing hazards abound.

Clearly, most stars even if orbited by Earth-like rocky planets, would be unable to sustain complex life. Only 10% of the Milky Way's stars reside in galactic habitable zone with chemical and environmental conditions suitable for the development of complex Earth-type life.

## 4. Evolution of Probable Habitable Planets and Satellites

Lammer et.al. (2009) have proposed a very useful classification of for planets and the evolution of life, and have demarcated four habitat types:

Class I habitats represent planets or moons where stellar and geophysical conditions allow Earth-like life to take root and which would enable complex multi-cellular life forms to evolve

Class II habitats include planets where life, similar to life on class I type planets, may appear and flourish. However, due to different stellar and geophysical conditions, life will evolve differently than on Earth, or it may flourish briefly and then die out. The planets under this category gradually become more like Venus or Mars-type worlds where complex life forms are unable to survive.

Class III habitats are planetary bodies with subsurface water layers or subsurface oceans which interact directly with a silicate-rich core (e.g. Europa). Presumably, microbes or simple multi-cellular eukaryotes may be able to survive in these environments.

Class IV habitats possibly have liquid water layers between two ice layers or liquid above ice (e.g. Ganymede, Callisto, Enceladus, and Titan-lakes). Life that evolves on Class IV planets or moons may include extremophiles, or forms which are not-carbon based or unlike those of Earth (Istock 2009; Naganuma and Sekine 2009;Schulze-Makuch 2009).

### 4.1 Class I Habitable Planets

Class I habitable planets are likely to be found orbiting around G-type stars and K and F-types with masses close to G- stars. These are the most likely candidates for harbouring complex multi-cellular surface life forms. The stellar and planetary geophysical conditions are such that an evolving atmosphere and watery environment can be maintained - perfect for the evolution of complex life.

Earth is an ideal example of planetary habitat supporting life under this category. NASA's Kepler Mission defines Earth-type planets to be those that have mass between 0.5 and 2.0 times Earth's mass, or those having between 0.8 and 1.3 times Earth's radius. Doppler studies and microlensing have uncovered a population of planets with masses 1.9-10 times the Earth's mass called super-Earth (Mayor et al. 2009 ; Beaulieu et al. 2006).

In December 2011, NASA's Kepler Mission announced discovery of first planet in the habitable zone of a Sun-like star. The exoplanet, 'Kepler-22b' is 2.4 times the size of Earth with an orbital period of 290 days, and has surface temperature similar to Earth's. It is closer to its parent star than Earth is to Sun.

Earth is in middle of our solar system's habitable zone with Mars and Venus on either side. It is ideally positioned within the solar system for life to evolve and diversify. It has liquid water, a breathable atmosphere, and a suitable amount of sunshine. If Earth were a little closer to the Sun, it might be cloaked in a thick veil of greenhouse gases, as is the case with Venus; a little farther out and it would become like cold arid Mars.

NASA's Kepler Mission is capable of detecting stars' dimming as small as one part in ten thousand of their normal brightness, providing sensitivity to detect planets even smaller than Earth. Then, NASA's Space Interferometry Mission, scheduled for launch in 2014, should be able to detect, image, and spectroscopically analyse the landscapes, oceans, and atmospheres of the terrestrial planets around several nearby stars (Marcy 2009).The European Space Agency(ESA)'s proposed Darwin probe and NASA's Terrestrial Planet Finder (TPF) probes will also have the capability to image Earth-like planets directly through development of space interferometer wherein combination of signals from constellation of individual telescopes will mimic a much larger telescope, needed to provide substantially enhanced resolution of the gap between the planet and the star. Darwin is also designed to detect and perform spectroscopic analysis of the atmosphere of Earth-like planets at mid-IR wavelengths (6-20 μm) for evidence of habitability and life (Cockell et al. 2008). The spectroscopic analysis would help establish presence of biomarker gases such as $O_2$ and its photolytic product,$O_3$, $H_2O$,$CO_2$, $N_2O$, and $CH_4$ on the atmosphere of probable habitable planets. The TPF series namely, Terrestrial Planet Finder Coronograph or Occulter (TPF-C/O) and Terrestrial Planet Finder Interferometer (TPF-I), operating in the visible/IR, too would be providing spectroscopic information on the existing atmosphere of Earth-like habitable planets (Alexander 2006).

### 4.2 Class II Habitable Planets

Class II habitats include terrestrial planets, which evolve differently from Earth. Class II planets are believed to orbit within Habitable Zones of low mass M and K-type stars. However, they are located either at the very edge of the habitable zone or very close to their suns so that their atmosphere - magnetosphere environments experience extreme exposure to stellar radiation and plasmas. Thus, they are likely to lose their atmosphere and oceans, which would make it difficult for life as we know it to evolve.

However, it is possible that even under these conditions, there may be niches where life may flourish, such as deep beneath the soil, particularly if early in its history the planet had oceans of water. Consider Venus or early Mars, both of which are believed to have had oceans of water early in the history. On the surface of Venus, liquid water is not supposed to exist because surface temperature (480ºC) is above critical temperature 374ºC for pure $H_2O$.

Exoplanets, whose surface is below the critical temperature, might host stable liquid water conditions under specific pressure conditions. Venus-like exoplanets with surface temperatures below 150ºC are likely candidates for sustaining life (Cockell 1999).

Mars is another example of terrestrial planet under this category. Today, Mars is a dry, frozen desert that cannot sustain life on its surface. During 4 - 4.5 Gya, Mars may have had an atmosphere thick enough to maintain liquid water on the surface (Kulikov et al. 2007). Large impactors evaporated the Martian atmosphere in planet's early history, and low gravity of the planet was not able to retain the gas in hot plumes created by these impactors (Melosh and Vickery 1999). Yet another factor is the loss of
its magnetic field, exposing the planet to the Sun's solar winds which may have blown off the atmosphere and oceans of Mars (Joseph 2009).

Liquid water cannot exist on the surface of Mars on account of its low atmospheric pressure of about 0.6 kPa compared to Earth's 101.3 kPa (Chambers 1999). However, water, and life, could exist just inches below the surface. NASA's Phoenix Mars Lander, which landed on the Mars Arctic Plain in May 2008, however, confirmed presence of frozen water near the surface. In June 2000, evidence for subsurface water was discovered in the form of flood-like gullies. In March 2004, NASA announced that its rover, 'Opportunity' had gathered additional evidence that Mars, in the ancient past, was a wet planet. Mars Express orbiter had directly detected huge reserves of water ice at Mars' south pole in January 2004.

Trace amounts of methane were also discovered in the Martian atmosphere in 2003 and verified subsequently in 2004 (Krasnopolskya et al. 2004). Presence of methane on Mars is very intriguing,since it is an unstable gas. Its presence indicates that there must be an active source on the planet in order to maintain such levels in the atmosphere. Mars produces around 270 tonnes of methane annually (Vladimir 2005), but asteroid impacts account for only 0.8% of the total methane production.Lack of current volcanism and hydrothermal activity or hotspots are not favourable for geologic methane. The existence of life in forms of microorganisms such as 'methanogens' to be the most likely source. If microscopic Martian microbial life is producing methane, these microbes probably reside far below the surface, where it is still warm enough for liquid water to exist.

In February 2009, the European Space Agency (ESA) announced discovery of the smallest super-Earth yet measured orbiting the star CoRoT-7. Although the planet orbits its host star once every 20 hours at a distance of < 0.02 AU, its diameter is estimated to be ~1.7 times that of the Earth. Due to its extreme closeness to its parent star, it is believed to have a molten surface temperature of 1000-1500ºC.

Charbonneau et al. (2009) have recently reported discovery of an exoplanet only 2.68 times larger than the Earth. The discovered super-Earth, 'GJ 1214b',with a mass 6.55 times Earth's mass, is speculated to contain water body larger than Earth, surrounding an inner core of iron and nickel, and an outer mantle of silicate rock. It is also believed to have an atmosphere of hydrogen and helium. The planet orbits a small and faint M-type star, 'GJ 1214' only 13 parsecs away from the Earth.

**4.3 Class III Habitable Planets**

Class III habitats, are planetary bodies with subsurface oceans in contact with silicate core. The Jovian moon, Europa is an excellent candidate for Class III status. The astrobiological potential and possible habitability of class III habitats rest on the presence of liquid water an adequate energy source to sustain the necessary metabolic reaction. Europa's unlit interior is considered to be one of the most likely locations for extant extraterrestrial life in our solar system (Schulze 2001).

The magnetic field data from the Galileo showed that Europa has an induced magnetic field through interaction with Jupiter's. This raises the possibility of a subsurface salty liquid water ocean, being kept warm by tidally - generated heat (Greenberg 2005). Life in a subsurface, Europa's ocean could possibly be similar to microbial life in the deep oceans of Earth as can be found in hydrothermic vents. Anaerobic chemosynthetic bacteria that inhabit hydrothermic ecosystems, or those of Antarctic subglacial Lake Vostak, provide a possible model for life in Europa's ocean.

**4.4 Class IV Habitable Planets**

Under Lammer et al.(2009)'s proposed classification, Class IV habitats include those planets and planetary satellites which have subsurface water oceans or water reservoirs that do not interact with a typical silicate bearing sea-floor like silicate-core on Europa. Subsurface oceans are speculated to exist between different ice-layers at Jupiter's Ganymede and Callisto, and Saturn's Enceladus, and Titan( Naganume and Sekine 2009; Schulze-Makuch 2009).

Findings of the Galileo space probes suggest that Ganymede, the largest satellite in the solar system is surprisingly Earth-like in some ways, including possessing a magnetic field and probably a molten core. It is composed primarily of silicate rock and water ice. Ganymede is likely to be fully differentiated with an iron sulfide and iron-rich liquid core. A saltwater ocean is believed to exist nearly 200km below Ganymede's surface (McCord et.al. 2001).

The microbial ecosystems on Earth that do not rely on sunlight, oxygen or organics produced at the surface provides analogues for possible ecologies on Enceladus. There is quite a possibility that the liquid water environment existing beneath the south polar cap on Enceladus may be conducive to life. NASA's Cassini spacecraft has shown that Saturn's satellite - Enceladus possesses active plumes and jets whose source may include pockets of liquid water near the surface or a pressurized subsurface liquid water reservoir. The Ion Neutral Mass Spectrometer (INMS) instrument onboard Cassini has found non-condensable volatile species (e.g. $N_2$, $CO_2$, $CH_4$) in jet-like plumes over Enceladus' geologically active south polar terrain (Waite et.al. 2007). If $N_2$ is present, it may reflect thermal decomposition of ammonia associated with the subsurface liquid reservoir, suggesting that water is coming into direct contact with very hot rocks; providing a source of heat as well as mineral sources for catalyzing reactions. If this scenario really exists on Enceladus, it is possible to speculate that the necessary ingredients may be present for the abiogenic creation of life through chemoautotrophic pathways. Ruling out for the moment, panspermia which is the most likely explanation for the presence of life on this body or other moons or planets (Joseph 2009; Wickramasinghe et al., 2009), it is reasonable to speculate that life on Enceladus and on Earth may have evolved around deep-sea hydrothermal vents. In this model, life on Earth began in deep sea hot springs where "arqueobacteria" extracted their energy from $H_2S$ and other Fe-compounds that billow out of the sea-floor.

Cassini-Huygen discoveries are indicative of Titan's potential for harbouring the ingredients necessary for sustaining life (Naganuma and Sekine 2009). These discoveries reveal that Titan is rich in organics, and contains a vast subsurface ocean with sufficient energy sources to drive chemical evolution (Coustenis and Taylor 2008).Recent models of Titan's interior, including thermal evolution simulation, predict that the satellite may have an ice crust between 50 and 150 km thick, lying atop a liquid water ocean a couple of hundred kilometre deep, with some amount (~10%) of ammonia dissolved in it, acting as an antifreeze. Beneath lies a layer of high-pressure ice. Cassini's measurement of a small asynchronicity in Titan's rotation is probably explained by separation of the crust from the deeper interior by a liquid layer (Lorenz 2008).

Titan's organic inventory is impressive and carbon-bearing compounds are widespread across the surface in the form of lakes, seas, dunes, and probably sedimentary deltas at the mouths of channels (Naganuma and Sekine 2009). Methane on Titan seems to play the role of water on the Earth. Earth has a hydrological cycle based on water, and Titan has a cycle based on methane. Cassini Spacecraft's close flybys have revealed presence of surface features on Titan similar to terrestrial lakes and seas. NASA scientists have identified a large lake near Titan's south pole which is filled largely with ethane, and smattering of methane, nitrogen, and other hydrocarbons.

Analogies can also be made between the current organic chemistry on Titan and the chemistry which was active on the primitive Earth. Titan is the only planetary body in our solar system other than Earth that is known to have liquid bodies on its surface. Several of the organic processes, which are occurring today on Titan, form some of the organic compounds which are considered as key molecules in terrestrial pre-biotic synthesis, such as hydrogen cyanide (HCN), cyanoacyetylene ($HC_3N$), and cyanogen ($C_2N_2$). Presence of nitrogen (95%) methane, ethane, and other simpler hydrocarbons in Titan's atmosphere makes it the most favourable atmosphere for prebiotic synthesis. Ethane and other hydrocarbons are simply products from atmospheric chemistry caused by the breakdown of methane by sunlight.

# 5. Probability of Life on Habitable Moons of Exoplanets

The current ground-based telescopes could detect an Earth-like exomoon in the habitable zone around a Neptune-like exoplenet (Kipping 2008). According to one study, habitable exomoons are detectable
up to ~100-200 parsecs away around early - K, M, and later-G dwarf stars by combining transit time variations (TTV) and transit duration variations (TDV) observed in respect of the transiting planet, down to 0.2 times Earth's mass using Kepler-class photometry (Kipping et al. 2009 ; Campanella 2009). The underlying principle is that an exomoon is supposed to induce TTV and TDV on the host planet. While TTV is caused by the position of the planet oscillating around the common centre-of-gravity (barycenter) of the planet-moon system, TDV is caused by the apparent velocity of the planet increasing and decreasing as it moves around the barycenter of the planet-moon system. These effects are predicted to exhibit a $\pi/2$ phase difference providing a unique detection signature for an exomoon.

In our solar system, it is observed that the larger a gas giant, greater the total mass of its satellites. Going by this norm, extrasolar giants-more massive than Jupiter may have moons as large as Mars or even Earth (Le Page 2008). One important factor determining a moon's suitability for habitability of life is the stability of its orbit, which can be disrupted by the close proximity to its Sun. Computer simulations suggest that a moon with an orbital period < 45-60 days will remain safely bound to a massive gas giant or brown dwarf that orbits 1AU from a Sun-like star. The major moons of our own solar system's gas giants all have orbital periods well within this range, between 1.7 and 16 days. At the lower end of the range, the orbit will still stay outside the Roche limit where a moon would be sheared apart by tidal forces. Thus, giant planets and brown dwarfs in a star's habitable zone indeed seem likely to have large moons in stable orbits giving rise to possibility of traces of life being detected thereon.

Presence of atmosphere is another prerequisite for development of life-system on a moon. In order to have enough gravity to hold on to an atmosphere, an exomoon is required to be larger than Earth's own airless Moon, which has a mass 0.012 times that of Earth. For a body with Mars-like density and an Earth-like atmospheric temperature profile, calculations show that the mass of the habitable moon must be at least 7% of Earth's to facilitate retention of most of its atmosphere for 4.6 billion years (Earth's current age). One of the criteria for retention of atmosphere on any moon is the geologic activity needed to keep the carbonate-silicate weathering cycle going, which regulates the global atmospheric temperature on geologic time scales. Without this self-regulating cycle, an otherwise habitable world will fall into perpetual ice age, much as Mars is today.

Length of a moon's day is yet another significant factor responsible for habitability of life on a moon. Computer simulations have demonstrated that any large moon orbiting a giant planet or brown dwarf becomes locked into a synchronous rotation - one side of the moon always facing the planet - within a few hundred million years. This happened also with Earth's own Moon and many other large moons in the solar system. Assuming that large moons typically have orbital periods of 2-16 days, any potentially habitable moon would have a day several times larger than Earth's. Simple calculations by Stephen Dole of the Rand Corporation in 1960's showed that a body with an Earth-like atmosphere would become uninhabitable when the period of rotation exceeds 4 days, due to large swings in surface temperature (Le Page 2008).

Eccentricity is also an important parameter in deciding suitability of a moon for habitability of life. Though a number of recently discovered exoplanets have mean orbital distance that lies within the habitable zones of their stars, most of them have eccentric orbits that would not be conducive to habitability of life on the exoplanets as well as their moons on account of large variations in the amount of sunlight reaching them. The mean insolation (measure of solar radiation energy in a given time) of the planet orbiting 16 Cygni B, for example, is about half that of the Earth - but in reality, ranges from 20% to 260% of the sunlight on Earth because of the planet's eccentric orbit. Living entities, if any, present on the planets and their moons would have a hard time being repeatedly deep-frozen and oven-roasted.

## 6. Panspermia

Most scientists subscribe to the view that life began through as yet unknown processes which yielded protocells possessing amino acids and nucleotides that were sparked with life (Menor Salván 2009; Sidharth 2009). It is reasonable to assume that if life could arise on Earth through abiogenesis, then the same events can take place on other worlds if they possess the same chemistry. If so, life should be widespread, particularly microbial life, in the universe. More complex life is another question. However, if those advocating panspermia are correct, life only had to arise once, and could then be dispersed from planet to planet and solar system to solar system.

Joseph (2009) presents evidence indicating that periodic and powerful solar winds blow portions of the Earth's atmosphere into space, along with airborne microbes. Hoyle and Wickramasinghe (2000) maintain that life-carrying dust and debris from Earth is inevitably distributed throughout our solar system, and even on a galaxy-wide scale. Wickramasinghe and colleagues (2009) also theorize that comets hitting the Earth could eject material containing life into space only to land on other moons and planets.

In support of these intriguing possibilities is the accumulating evidence that microbes can survive the ejection from and the landing onto the surface of a planet, and a journey through space (Joseph 2009; Wickramasinghe et al. 2009). Microbes can also remain in dormant state for hundreds of millions of year before returning to life (Vreeland et al. 2000; Gilichinsky 2002) - findings which appear to support the advocates of panspermia.

If Wickramasinghe et al., (2009), Joseph (2009) and the 'seeds of life' advocates are correct, life could have begun on Earth, and microbial life could have been and could continue to be deposited on every moon and planet in our solar system. Survival of those forms of life would depend on whether the planet or moon is a Class I, II, III, or IV. It is important to stress, however, that the advocates of panspermia do not believe life began on Earth but has a source much older than our planet (Joseph 2000, 2009; Sharov 2009; Wickramasinghe et al. 2009).

## 7. Concluding Remarks

Discovery of over 2,800 exoplanets orbiting distant stars supports the likelihood that most Sun-like stars may have planets around them. Astronomers at the Harvard-Smithsonian Center for Astrophysics (CfA) studying data from the Kepler Space Mission estimate that 17% stars in the Milky Way galaxy have planets about the size of Earth, meaning that there is a minimum of 17 billion Earth-size planets in our home galaxy (Kepler Astrophysics News 2013). Kepler could confirm discovery of 135 Earth-size planets and spotted nearly 900 extrasolar planets before it went out of operation due to snags with its stabilizing system in May 2013.

Discovery of wide range of extremophiles in inhospitable environments on Earth has raised possibility that primitive microbial life may survive on several planets and moons in our solar system as well as on countless exoplanets in other solar systems. Moreover, search for life could not be limited to search for Earth-like features alone. We need to expand the boundaries of our Earth-centric concept of life. Life on other worlds may have a completely different chemistry, and may not even possess a genetic code. It might be based on molecular structures substantially different from those we know, and hence could be unrecognizable compared with life observed on Earth, nor could they be detectable by telescopes and space probes designed to detect terrestrial biomarkers either. Life forms based on silicon, ammonia, and sulfur are among those that may have evolved on other worlds (Pabulo 2010).

Presently, we have very limited information about the location of either the constituents of building blocks of life or the building blocks themselves in our solar neighbourhood. It still remains far from clear as to what constitutes life or how life was first established. Amino acids, supposedly the building blocks of life, are definitely not living entities. They are complex organic molecules present in all living organisms facilitating production of protein, which is so vital for genetic replication. Their existence barely demonstrates that a complex organic chemistry is at work. No laboratory experiment involving interactions among the known building blocks of life has produced any trace of life as yet. The mere presence of organic molecules or basic building blocks of life, viz. amino acids, purines, and pyrimidines in interstellar clouds, comets, and exoplanets does not ensure presence of life on them, though it certainly raises the possibility that they were biologically produced (Joseph, 2009 Wickramasinghe et al., 2000). If true, then life may be everywhere as claimed by the "seeds of life" advocates of panspermia.

At present, however, panspermia remains at best an intriguing possibility. Although there is no evidence to date that life can be created from non-life despite considerable efforts in well-funded laboratories, the origin of life, and the ease at which it may take root on other worlds, remain unanswered questions. Until these issues are resolved satisfactorily and the nature of life is better understood, we can only say with certainty that Earth is the lone habitat in the universe where life has achieved an authentic foothold.